\documentclass[12pt]{iopart}
\usepackage{graphicx}
\usepackage{xspace}
\newcommand{\pT}{p$_{T}$ \xspace}
\newcommand{\axi}{$\overline{\Xi}^+$ \xspace}

\newcommand{\aomega}{$\overline{\Omega}^+$ \xspace}
\newcommand{\alam}{$\overline{\Lambda}$ \xspace}
\newcommand{\sqrts}{$\sqrt{s_{_{NN}}}$ \xspace}

\begin{document}

\title[What's Interesting About Strangeness Production?]{What's Interesting About Strangeness
Production? - An Overview of Recent Results}

\author{Helen Caines
\footnote[3]{Correspondence address: helen.caines@yale.edu} }

\address{Yale University, Physics Department, WNSL, 272 Whitney Avenue,
New Haven, CT 06520, U.S.A.}

\begin{abstract}

\noindent In this paper I highlight a few selected topics on strange particle
production in heavy-ion collisions. By studying the yield and
spectra of strange particles we hope to gain understanding of
the conditions reached in, and the ensuing dynamics of, the systems
produced when ultra-relativistic heavy-ions are collided.

\end{abstract}


\section{Motivation}

The argument for studying strange particle production in heavy-ion
collisions as evidence of Quark-Gluon Plasma (QGP) formation is very
simple and was first suggested in 1982~\cite{Raf:StrangeSig}. The
idea relies on the difference in production rates of strange
particles in a hadron gas compared to strange quarks in a QGP. As
there are no strange quarks in the in-coming colliding nuclei and
strangeness is a conserved quantity each strange particle produced
must be accompanied by a corresponding particle containing an
anti-strange quark. In a hadron gas the energy threshold for strange
particle production is high. The creation of  a $\Lambda$ is predominantly
through the reaction $\pi$ + N $\rightarrow \Lambda$ + K,
with a threshold energy requirement of $\sim$530 MeV. While that
of an \alam requires E$_{thresh}$ $\sim$1420 MeV. Multi-strange particle creation not
only needs a large amount of energy but is also a multi-step reaction as
first a singly strange particle and then a multi-strange one must be
created. Should the system convert to one whose constituents are quarks
and gluons the situation
simplifies. The energy threshold, for strangeness production is now reduced to
$\sim$300 MeV, or
twice the strange quark mass (a quark/anti-quark pair must be
created). Thus strange quarks become much more abundant and upon
hadronization the relative density of (multi-)strange particles is
significantly enhanced over that resulting from a hadron gas. It
should also be noted that in a QGP system the gluonic degrees of
freedom dominate and the cross-section for gg$\rightarrow s\bar{s}$
is much larger than the cross-section for s$\bar{s}$ creation in a
hadronic gas. Thus, not only is creation energetically favourable but the
probability is larger. The greatest enhancement in yields is
expected to be for multi-strange anti-particles such as the
\aomega.

    Although these arguments are old and simple they are still
valid today. However, studies of strange particle production and
spectra are now made to investigate many other topics relevant to
heavy-ion collisions and the dynamics of the source produced.

\begin{figure*}[htbp]
\centering
  \includegraphics[width=0.5\textwidth]{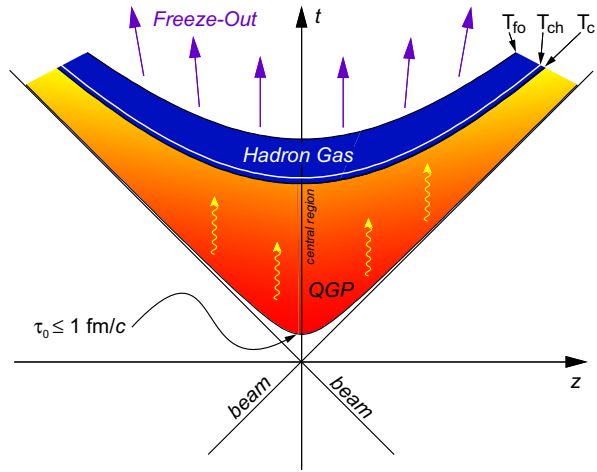}
  \caption{Schematic graph indicating the course of a heavy-ion collision, assuming
  the system passes through a QGP stage. See text for further details. }\label{Fig:LightCone}
\end{figure*}

    Figure \ref{Fig:LightCone} highlights the important evolution
sequences of a collision. The initial pre-equilibrium phase is
dominated by hard scatterings, by which we mean high
Q$^{2}$ collisions of  partons. The physics of this early time is probed by
high p$_{T}$ particles and jet phenomenology. Much progress has been made in
this area especially, through the use of strange particles to make
identified hadron measurements, and it was the topic of two
other sessions at this conference. Therefore, I do not cover high \pT physics
in this paper.

    We aim, in the early stages of the collision, to create a temperature
that exceeds T$_{c}$, the critical temperature at which a transition to
partonic degrees of freedom occurs. The excited region then
expands and cools and drops below T$_{c}$, when hadrons
 re-form. It is believed that there is a large amount of
re-scattering both in the partonic and hadronic phases and that the system
reaches chemical equilibrium. Soon after
T$_{c}$ the system passes through chemical freeze-out, T$_{ch}$. At
this point inelastic scatterings cease and the stable hadron ratios
are frozen in. Finally kinetic
freeze-out occurs when the system cools to below   T$_{fo}$ and elastic collisions also end.
After this the particles free stream to the detectors without any further interaction.

       After first briefly, in section \ref{Section:tech},
discussing the experimental and reconstruction techniques used to
identify strange particles I have organized this paper along the
time-line of a collision as shown schematically in
Fig.~\ref{Fig:LightCone}.  In section \ref{Section:BaryonNumber}, I use the strange hyperon
rapidity spectra and anti-particle/particle ratios to probe how incoming protons and neutrons
are transported from beam to mid-rapidity as a function of collision energy.

    The issue of chemical equilibrium is discussed in
section ~\ref{Section:Chemistry} and finally,  section
\ref{Section:Dynamics} describes how the \pT spectra are used to probe
the source dynamics at kinetic freeze-out.

\section{Techniques for Identification}\label{Section:tech}

Table~\ref{Table:WeakDecay} lists some of the weakly decaying strange
particles, their quark content, dominant decay modes and their
lifetimes. As can be seen many of these particles are either neutral
or their major decay channel contains a neutral ``daughter'' particle,
and they have lifetimes of only a few cm. Thus the majority of strange
particles are identified via their decay topologies. The exception
being the charged kaon, although this too is often reconstructed at
higher \pT through its distinctive kink topological decay.

\begin{table}[hb]
\begin{center}
\begin{tabular}{|c|c|c|c|}
  \hline
  Particle & Quark Content & Dominant Decay Mode & Lifetime (c$\tau$)\\
  \hline
  $K^{\pm}$ & ($u\bar{s}$, $\bar{u}s$)& $\mu^{\pm}$ + $\nu_{\mu}$& 3.7 m \\
   $K^{0}_{s}$ & ($d\bar{s}$+$\bar{d}s$)& $\pi^{+}$ + $\pi^{-}$ & 2.7 cm \\
   $\phi$ & $s\bar{s}$ & $K^{+}$ + $K^{-}$ & 44.6 fm \\
  $\Lambda$ & $uds$ & $p$ + $\pi^{-}$ & 7.9 cm \\
   $\Xi^{-}$ & $dss$& $\Lambda$ + $\pi^{-}$ & 4.9 cm \\
  $\Omega^{-}$ & $sss$ & $\Lambda$ + $K^{-}$ & 2.5 cm \\

  \hline
  \end{tabular}
   \end{center}
  \caption{ The major weak decaying strange particles, their quark
  content, dominant charged particle decay modes and lifetimes.}
  \label{Table:WeakDecay}

\end{table}

\subsection{Topological Reconstruction}

Once identified through their decay, strange particles
are selected using invariant mass analysis. The decay point of the mother particle
is  located via the secondary vertex of the decay. The decay of the
strange particles usually occurs before the tracking detectors'
active volumes, hence a visual identification of the vertex is
 not possible and reconstruction via combinatorial methods is employed.
 The decay topologies come in three distinct types and
are known by the pattern they are identified via: the ``V0", the
``Cascade", and the ``Kink". These methods are described below in turn.

\subsubsection{The ``V0"}
~

    This pattern is produced by the decay of a neutral
particle into two charged ``daughters". The neutral particle leaves
no ionization trail, but upon its decay a ``vee" appears in the
chamber as two oppositely charged tracks apparently appear from
nowhere.  Reconstruction of these decay modes generally
proceeds via the following process. All oppositely charged tracks
are paired and projected towards the primary collision
vertex. The two trajectories are compared to see if they appear to
cross at a point before the primary vertex. If so these two
tracks are considered candidates for a ``V0'' decay. The momentum
components  of the two charged daughters, at the now assumed decay
point, are then used, assuming of the daughter particles'
masses for the decay of interest, to calculate the  mother's
invariant mass using Eqn.~\ref{Eqn:InvMass}.  E$_{1}$ and
E$_{2}$ signify the two charged daughter energies and p$_{x1}$ etc,
their individual momentum components.
\begin{equation}\label{Eqn:InvMass}
    M_{invariant} = \sqrt{(E_{1} + E_{2})^{2} - p_{mother}^{2}}
\end{equation}
\begin{equation}
    p_{mother}^{2} = (p_{x1}+p_{x2})^{2} +  (p_{y1}+p_{y2})^{2} +
      (p_{z1}+p_{z2})^{2}
\end{equation}
If the identified tracks do indeed originate from the decay of the
desired mother the invariant mass calculation will result in the correct
mass, within  the momentum resolution of
the tracking detectors.

\subsubsection{The ``Cascade"}
~

    The ``Cascade'' decay is similar to the ``V0'' except there are now
 two decay vertices. Typically the mother decays first into a charged particle
and a neutral one, which subsequently decays further into two
charged daughters. Thus the decay creates a cascade of charged
particles.
 The reconstruction of the ``Cascade'' decay  occurs in a two
 step manner. First the decay vertex of the neutral daughter
  is reconstructed  using the ``V0'' technique just described, and the
  mass and momentum components of the particle are determined. This neutral
   particle is then combined with all appropriately charged tracks for the
 mother's decay mode and another secondary vertex is sought. Again the
 invariant mass is calculated assuming
 the masses and momenta of the decay products at the secondary
 vertex.

\begin{figure}[htbp]
 \begin{minipage}{0.46\linewidth}
    \begin{center}
  \includegraphics[height=\linewidth,angle=-90]{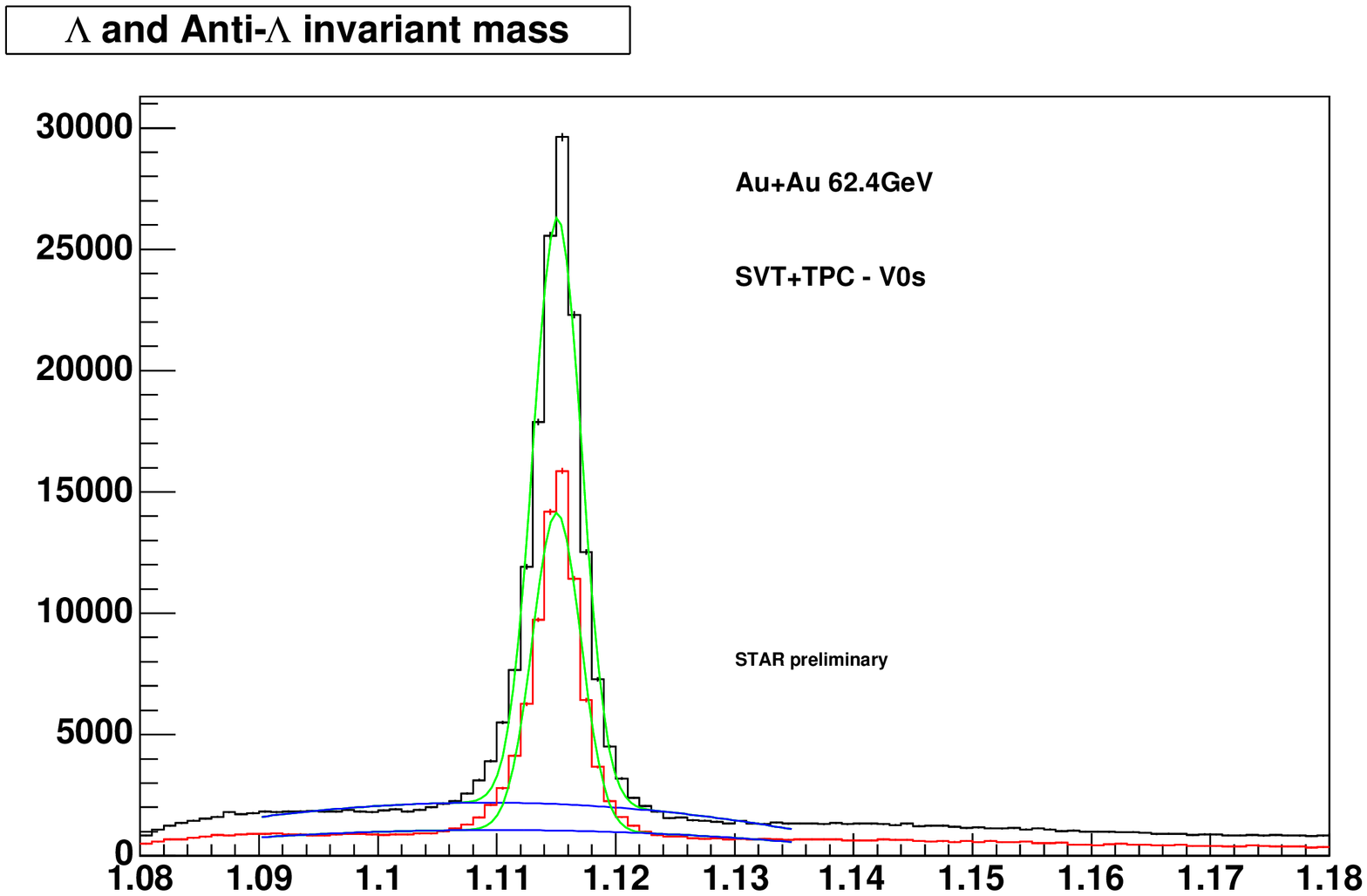}\\
  \caption{$\Lambda$ and \alam invariant mass peaks from Au+Au collisions at
  \sqrts= 62.4 GeV.}\label{Fig:InvMass}
  \end{center}
\end{minipage}
\hspace{1cm}
\begin{minipage}{0.46\linewidth}
 \begin{center}
  \includegraphics[width=\linewidth]{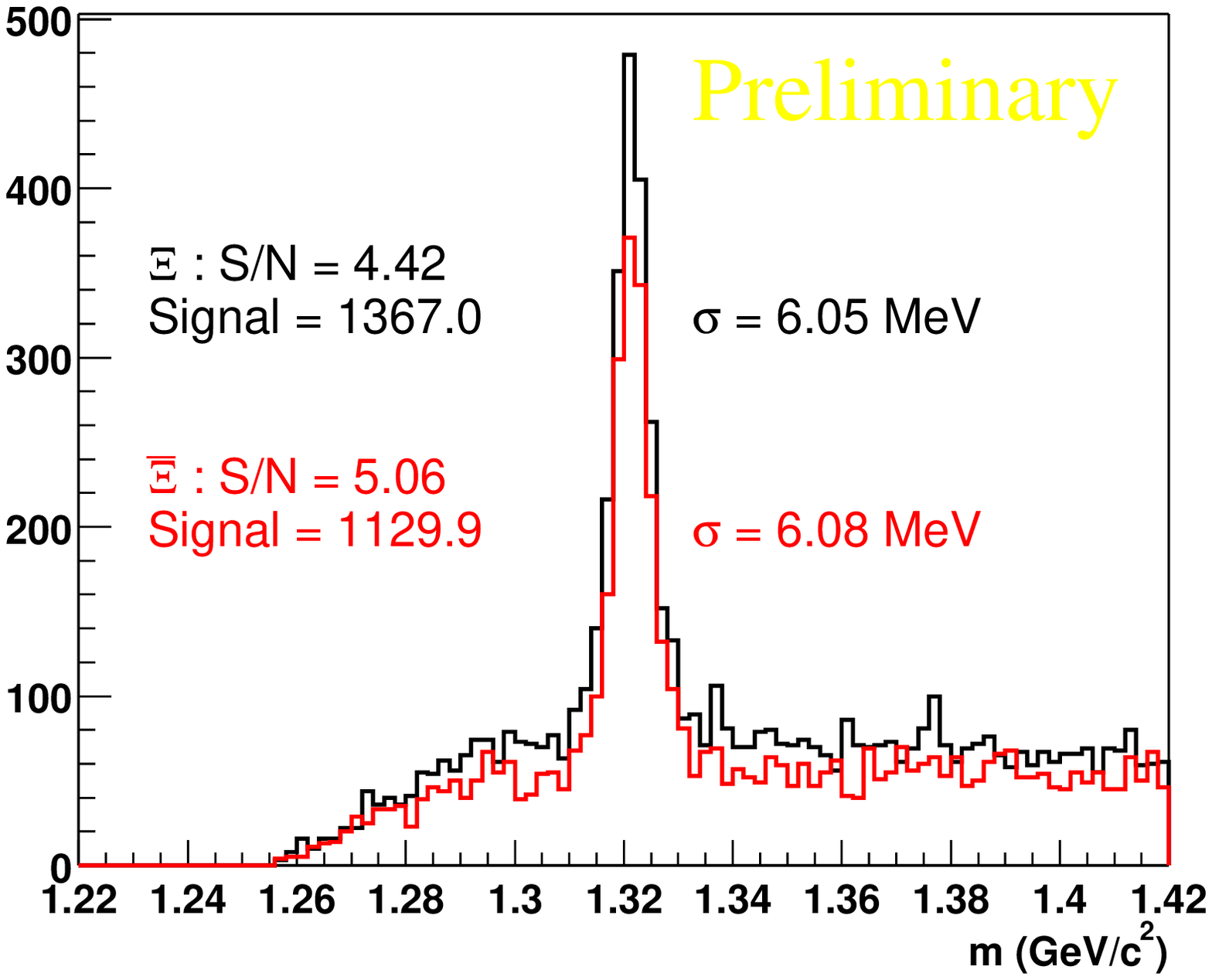}\\
  \vspace{-0.5cm}
  \caption{$\Xi^{-}$ and \axi invariant mass peaks from Au+Au collisions at
   \sqrts = 200 GeV.}\label{Fig:InvMass2}
   \end{center}
   \end{minipage}
\end{figure}

    In a heavy-ion collision, where many charged particles are
produced, these two techniques produce an enormous amount of random
combinatorial background candidates which must be eliminated. This
is done by applying various geometrical cuts to the each candidate.
 For instance, the parent should
appear to be consistent with a particle being emitted from the primary collision
point. The various daughter tracks should not have such
trajectories. Another common requirement is that the secondary vertex occur
more than a specified distance from the primary vertex. While this
eliminates many true decay vertices it commonly removes more
background than signal. As this cut is typically of several cm's, it is obvious that the efficiency
for strange particle reconstruction is low, of order a few percent.
Nevertheless these methods are very
successful at identifying clean signals as shown in the invariant mass plots for
$\Lambda$ and $\Xi$ from STAR  in
Fig.~\ref{Fig:InvMass} and Fig.~\ref{Fig:InvMass2}.

\subsubsection{The ``Kink''}
~

    The ``Kink'' is the result of a charged particle decaying into a stable
neutral particle and a charged one. This reconstruction technique is
only used when the lifetime of the parent is long and there is a
high probability of the parent leaving a signal in the tracking
chambers. Firstly all charged tracks appearing to emanate from the
primary vertex are considered. These tracks are studied to see
if their ionization paths terminate within the chamber. If such candidates are found
all tracks of the same charge, whose ionization paths appear to start
further away from the primary vertex than that of the now candidate
parent, are considered. The two tracks are projected towards one another
to see if they cross. If so this
pair of tracks are considered to be a decay candidate. An invariant
mass analysis is not possible with this technique as the neutral
particle in the decay is not identified, hence the charged track
appears to suddenly ``kink'' within the detector. As this technique is most commonly used to
identify charged kaon decays, the  main source of background
is the pion decay. Kinematics force the pion to decay with a
very small angle, so by insisting that the decay fall within
a certain window in decay angle, and applying other geometrical cuts,
the kaon can be cleanly identified.

     All these techniques while having low efficiencies have an
advantage over other direct particle identifications, such as dE/dx
and TOF, in that they can be applied over  large \pT ranges. The
limit is generally the available statistics which can always be
redressed through more beam time.

\section{Yields and Baryon Transport}\label{Section:BaryonNumber}

    It is still a matter of debate how baryon number is carried by
nucleons. It is however clear that baryon number is a conserved
quantity. By colliding nuclei at high energies we can study
how this baryon number becomes distributed across the whole
collision region and hope to gain further insight. At top RHIC energies the
beams are separated by over 10 units of rapidity. It is hard to comprehend how
anything massive can be transported over such a large rapidity gap,
so it was expected that the mid-rapidity region would be net-baryon free.

\subsection{ Net-Baryon Number at Mid-Rapidity}

\begin{figure}
 \centering
  \includegraphics[width=0.8\textwidth]{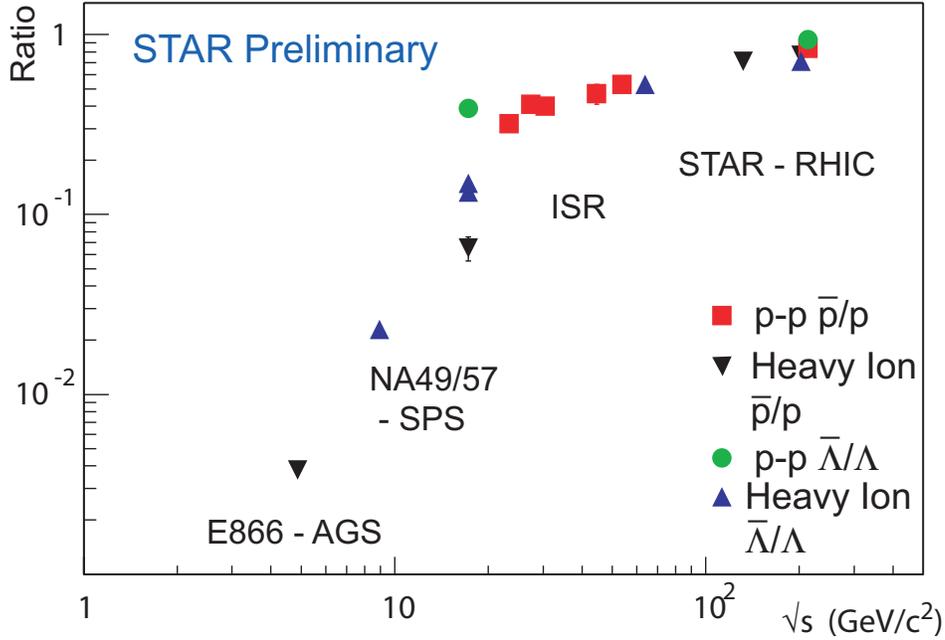}\\
  \caption{Anti-baryon/baryon ratios in $p+p$ and heavy-ion collisions vs. \sqrts.}\label{Fig:BbarB}
\end{figure}

    Fig.~\ref{Fig:BbarB} shows the anti-baryon to baryon ratios
($\bar{p}/p$ and \alam/$\Lambda$) for $p+p$ and heavy-ion collisions  as a function
of \sqrts~. It can be
seen that there is a smooth transition from baryon domination at low \sqrts to a near
net-baryon free region at \sqrts = 200 GeV. The new data from STAR at RHIC from
 Au+Au collisions as \sqrts= 62.4 GeV fits consistently into this trend.

    In RHIC \sqrts = 200 GeV collisions $\sim$1.6 TeV appears within
    $-1< \eta < 1$~\cite{PhobosWhitePaper}, this
means there is plenty of energy for the creation of strange
particles. As not all of this energy is concentrated in one rapidity
slice but spread over several units, it is interesting to
measure not just the mid-rapidity yields but their distributions.

\subsection{Rapidity Distributions}

\begin{figure}
\centering
  \includegraphics[width=0.9\textwidth]{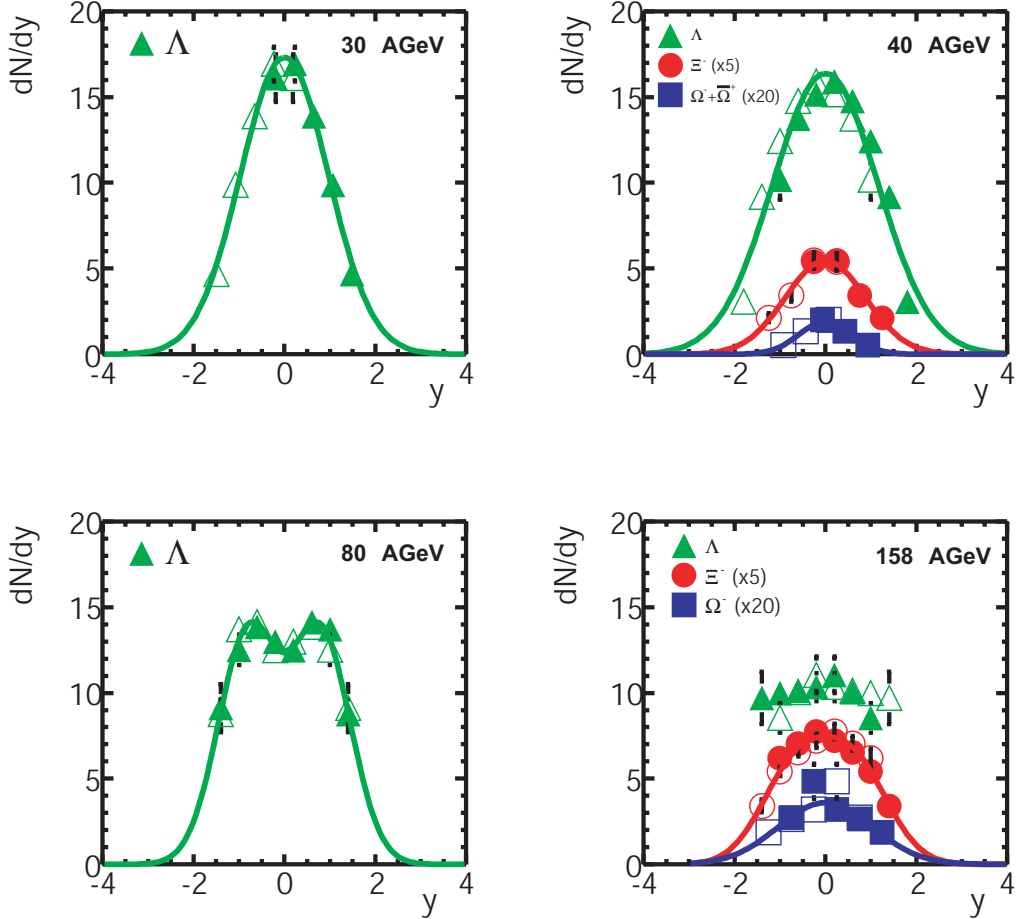}\\
  \caption{Energy evolution of the $\Lambda$, $\Xi$ and $\Omega$ rapidity
  distributions measured by NA49 at CERN SPS from \sqrts= 7.6, 8.8, 12.3, and 17.3 GeV.
  For clarity the $\Xi$ are scaled by a factor of 5 and the $\Omega$ by 20. The open
  symbols are a reflection around mid-rapidity. $\Lambda$ spectra are uncorrected for
  feed-down from multi-strange baryons. All spectra are preliminary except the $\Xi$
  at 17.7 GeV.}
  \label{Fig:LambdaRap}
\end{figure}

Fig.~\ref{Fig:LambdaRap} shows the rapidity
distributions of $\Lambda$  for four different energies at the CERN
SPS~\cite{Meurer}. It can  be seen that there is a distinct evolution in
shape as \sqrts increases. At \sqrts = 7.6 GeV the distribution is
Gaussian, however by \sqrts = 12.3 GeV two distinct peaks,
symmetric around mid-rapidity, are observed and at 17.3 GeV the
distribution is approximately flat between $\pm$ 1 in rapidity. The $\Xi$ and
$\Omega$ measured at 8.8 and 17.3 GeV show  a less distinct change,
although the Gaussian does broaden slightly in both cases. This is
further evidence of baryon transport from the beams. These results and others
of protons at RHIC energies~\cite{BrahmsNetProton}, show
clearly that on average the beam nucleons are shifted by $\sim$2
units of rapidity. At lower energies this corresponds to
transportation to mid-rapidity, or near complete stopping. At higher energies this results in
net baryon peaks away from mid-rapidity and a near net-baryon free
region at RHIC, as shown in Fig~\ref{Fig:BbarB}.

    It is also interesting to note that the mid-rapidity
ratios are flat as a function of centrality for all collision
energies suggesting that it is the collisions energy and not the
number of participants, $N_{part}$, that determines the fraction of the baryon
transport relative to pair production.

\subsection{ Mid-Rapidity Yields.}

\begin{figure}
 \centering
  \includegraphics[width=0.7\textwidth]{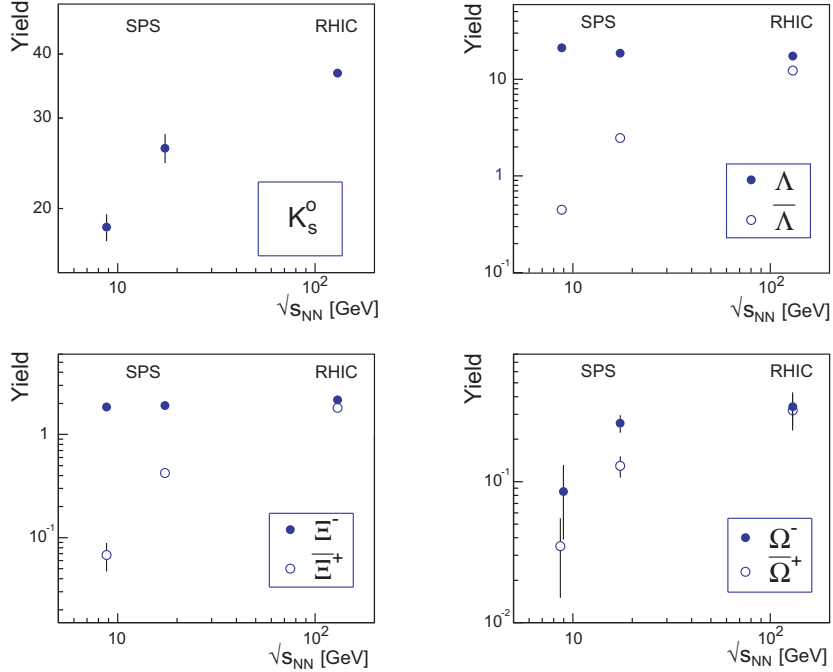}\\
  \caption{Mid-rapidity yields as a function of \sqrts. Data from
  \cite{STARHyperon} and \cite{NA57Hyperon}.}
  \label{Fig:Yields}
\end{figure}

 Several features can be seen in the \sqrts dependence of the
strange particle yields, shown in Fig.~\ref{Fig:Yields}. The most
striking is the difference in the \sqrts dependence of the baryons
and anti-baryons. This is once more a reflection of the changing
net-baryon number as the collision energy increases. The anti-baryon
and K$^{0}_{s}$ yields increase smoothly as the available energy
increases. The $\Lambda$ and $\Xi$ yields stay almost constant over
an order of magnitude increase in energy. This is an interesting
collusion of the decrease in baryon number, which causes a reduction
in the yields, being counteracted by the increase in available
energy.

\subsection{Kaon Ratios}

\begin{figure}
 \centering
  \includegraphics[width=0.5\textwidth]{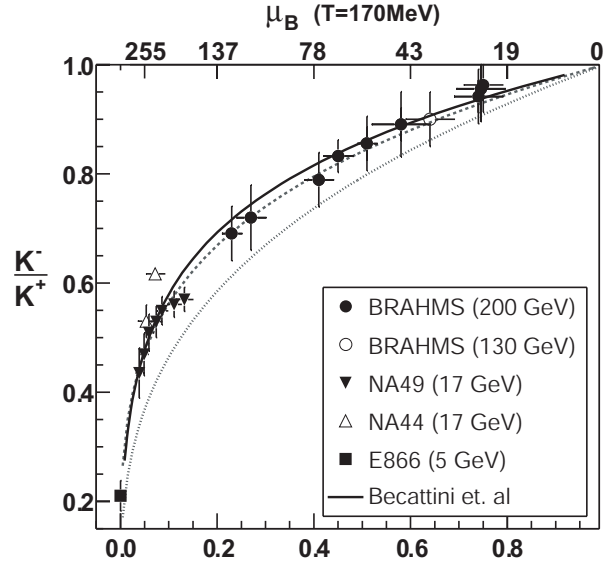}\\
  \caption{K$^{-}$/K$^{+}$ ratio versus $\bar{p}$/p for various collision energies and
  rapidities. Data from BRAHMS~\cite{BrahmsKRap}, NA49~\cite{NA49K+}, NA44~\cite{NA44K+} and
  E866~\cite{E866K+}. The dashed line shows K$^{-}$/K$^{+}$
    = ($\bar{p}$/p)$^{1/4}$, the dotted K$^{-}$/K$^{+}$= ($\bar{p}$/p)$^{1/3}$ and the
    solid the prediction from the statistical model described in
    \cite{Becattini}.}
   \label{Fig:BrahmsRap}
\end{figure}

    The quark contents of the $K^{+}$ and $K^{-}$ mesons are $\bar{s}u$
and $\bar{u}s$ respectively. Thus the $K^{-}$/$K^{+}$ ratio which
effectively =$\bar{u}/u$ as $\bar{s}/s$==1, also reveals information
about the systems' baryon content,
despite the Kaon being a meson. Fig.~\ref{Fig:BrahmsRap} shows the
correlation between the $K^{-}$/$K^{+}$ and the $\bar{p}/p$ ratios
for various beam energies and rapidities. As expected the two
ratios show a smooth correlation, again indicating a falling
net-baryon number with increasing collision energy. This correlation
is well represented by the power law function $K^{-}$/$K^{+}$ =
($\bar{p}/p$)$^{1/4}$, shown as the dashed curve, rather that the
function expected from a thermal interpretation with vanishing
strange quark potential of  $K^{-}$/$K^{+}$=
($\bar{p}/p$)$^{1/3}$, the dotted curve. This deviation might
be expected as the different measurements represent different
rapidity regions. However the solid curve shows the prediction from a
statistical model\cite{Becattini} assuming T = 170 MeV and it shows
good agreement. At RHIC energies the calculation show that the
chemical potential, $\mu_{B}$, is $\approx $ 130 MeV at y=3 compared to $
\approx $ 25 MeV at y=0. The basic premises of statistical models are
described in section~\ref{Section:Chemistry} below. The overlap of
the CERN measurements, taken at mid-rapidity, to those of the BRAHMS
measurements at forward rapidities should also be noted. This
suggests that, chemically at least, the medium created at
mid-rapidity at CERN occurs in the RHIC forward rapidities. This suggests the
possibility of studying
many different chemical environments by sitting at one collision
energy and merely altering the rapidity region.

\section{Chemistry}\label{Section:Chemistry}

Determining the chemistry of the particles
emitted from the collision region can tell us a great amount about the
source created. A common way to do this is using a statistical
hadronic model.

\subsection{Statistical Hadronic Models}

    A vast amount of work has been done implementing these models
to aid our understanding of heavy-ion collisions.
This discussion is not intended to be exhaustive but to give an
overview of the most salient points. Further details can be obtained
from ~\cite{StatModels} and references therein.

     The most important features of statistical models are
that they assume a thermally and chemically equilibrated system at
chemical freeze-out. They make no assumptions about how the system
arrived in such a state, or how long it exists in such fashion. They
also assume that the system consists of non-interacting hadrons and
resonances. Given these assumptions the number density of a given
particle, i, with mass m$_{i}$, momentum p, energy E$_{i}$, baryon
number B$_{i}$ and strangeness S$_{i}$, can be calculated for a
Grand Canonical ensemble where g is the spin degeneracy factor via
Eqn.~\ref{Eqn:StatModel}

\begin{equation}\label{Eqn:StatModel}
   N_{i} = \frac{g_{i}}{2 \pi^{2}} \int_{0}^{\infty} \frac{p^{2} dp}{\exp[(E_{i} -
    \mu_{B}B_{i} - \mu_{s}S_{i} )/(T_{ch}\pm 1)]}
\end{equation}

\noindent for a given chemical freeze-out temperature, T$_{ch}$,
baryo-chemical potential, $\mu_{B}$, and strangeness potential,
$\mu_{s}$. The conservation laws of baryon number, strangeness and
isospin have to be observed. In general these models are used to
determine T$_{ch}$, $\mu_{B}$, and $\mu_{s}$ of a given data set by
comparing ratios of different particles measured in the experiment
to those calculated via the model. The temperature and chemical
potentials are varied until a minimum in the comparison of the model
to data is achieved. To obtain a good description of the data as
many resonances as possible must be included. A study
of Eqn.~\ref{Eqn:StatModel} shows that different particle ratios
are sensitive in various degrees to T$_{ch}$, $\mu_{B}$ and $\mu_{s}$. For
instance anti-baryon to baryon ratios are highly sensitive to
$\mu_{B}$ but virtually insensitive to T$_{ch}$; the reverse being
true of baryon to pion ratios. As the system is believed to be short lived
there may not be time to fully saturate the strangeness content.
What strangeness there is can be evenly distributed through the
system however so the system can be thought of as in equilibrium, but the
strangeness phase space will not be saturated.  This non-saturation is often
accounted for in the models by the factor, $\gamma_{s}$, where $\gamma_{s}$ $\leq$ 1.

\begin{figure}[htbp]
 \begin{minipage}{0.46\linewidth}
    \begin{center}
     \includegraphics[width=\textwidth]{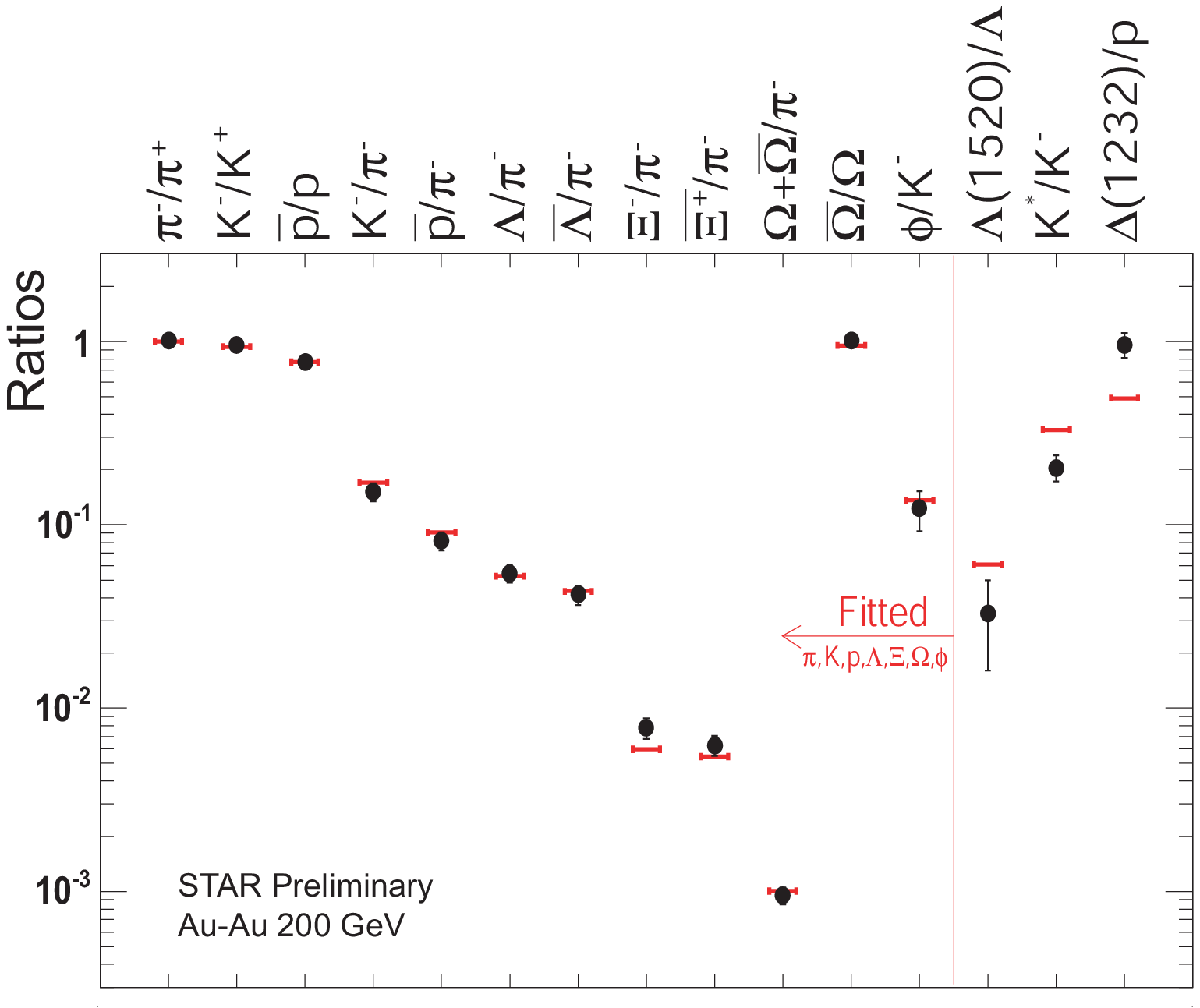}\\
  \caption{Measured particle ratios (symbols) and statistical model calculations (lines)
  \cite{PBMStat} to the 5$\%$ most central Au+Au data at \sqrts = 200 GeV. Errors shown are
  systematic.
  The measured yields of $\pi$, K, p, $\Lambda$, $\Xi$, $\Omega$,
  and $\phi$ are used in the fit. }\label{Fig:ChemFit}
  \end{center}
\end{minipage}
\hspace{1cm}
\begin{minipage}{0.46\linewidth}
 \begin{center}
   \includegraphics[width=\textwidth]{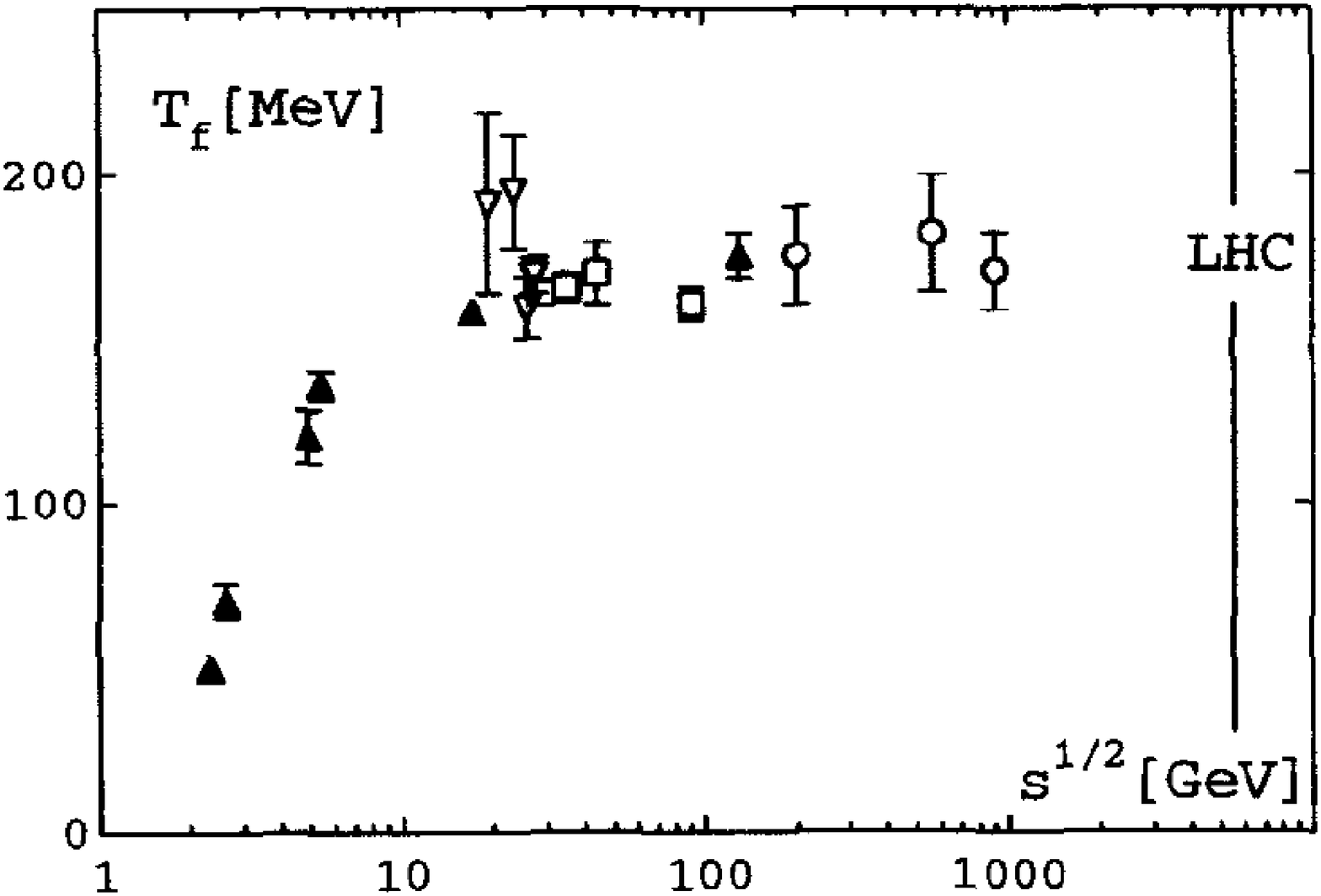}\\
  \caption{The calculated T$_{ch}$ as a function of collision energy, for e$^{+}$e$^{-}$ (squares),
   pp (open triangles) $\bar{p}+p$ (circles) and A+A (closed triangles). Data from \cite{Satz}.}
   \label{Fig:TchSystematics}
   \end{center}
   \end{minipage}
\end{figure}

     Fig.~\ref{Fig:ChemFit} shows the results of such a fit using the
statistical model described in \cite{PBMStat}. It can be seen that a
good representation of the data is possible, except for the short lived resonances. A
significant fraction of the particles are in
resonant states at chemical freeze-out thus it is vital that they are included in thermal
model descriptions despite this apparent failure to describe their yields. It is believed
 that the discrepancy between the calculation and measurements results from the measurable
 resonance yields  being altered after chemical freeze-out due to re-scattering and/or
 regeneration  in the hadronic phase. Neither re-scattering or re-generation alters the
 chemistry of the system they merely affect the measurable resonance signals.  See
\cite{Markert:HotQuarks} in these proceedings for more discussion
discussion of this effect. The results of the fit give T$_{ch}$ = 160 $\pm$ 5 MeV,
$\mu_{B}$ = 24 $\pm$ 4 MeV, $\mu_{s}$ = 1.4 $\pm$ 1.6 MeV, and $\gamma_{s}$
=  0.99 $\pm$ 0.07. This fit suggests that the Au+Au system at RHIC is very close to complete
strangeness saturation with near zero baryon and strangeness chemical potentials. Equally successful
fits are obtained at lower energies and Fig.~\ref{Fig:TchSystematics} shows the resulting T$_{ch}$
as a function of \sqrts. There is an apparent limiting
temperature reached, which is very close to the critical temperature, T$_{c}$ of 170 MeV,  predicted
 from Lattice QCD calulations\cite{LQCDTemp}.

     It is surprising however to see that statistical models appear to work even
for elementary collisions, see Fig.~\ref{Fig:TchSystematics}, so we should
treat the results with caution. To understand how this could be possible it should be remembered
 that all these data sets represent the yields of particles from an event ensemble average.
 Therefore, it is possible that the fits do not represent true temperatures and chemical
  potentials but are instead the Lagrange multipliers that will result from a fit to any
  statistically sampled data set. These fits will only have physical meaning if each individual
event can be thought of as a  statistically independent system. We therefore need to establish at
what collision energy, if any, state occurs and thus phase space considerations become
important.

\subsection{Phase Space Considerations}

\begin{figure}
\begin{minipage}{0.46\linewidth}
    \begin{center}
  \includegraphics[width=0.9\textwidth]{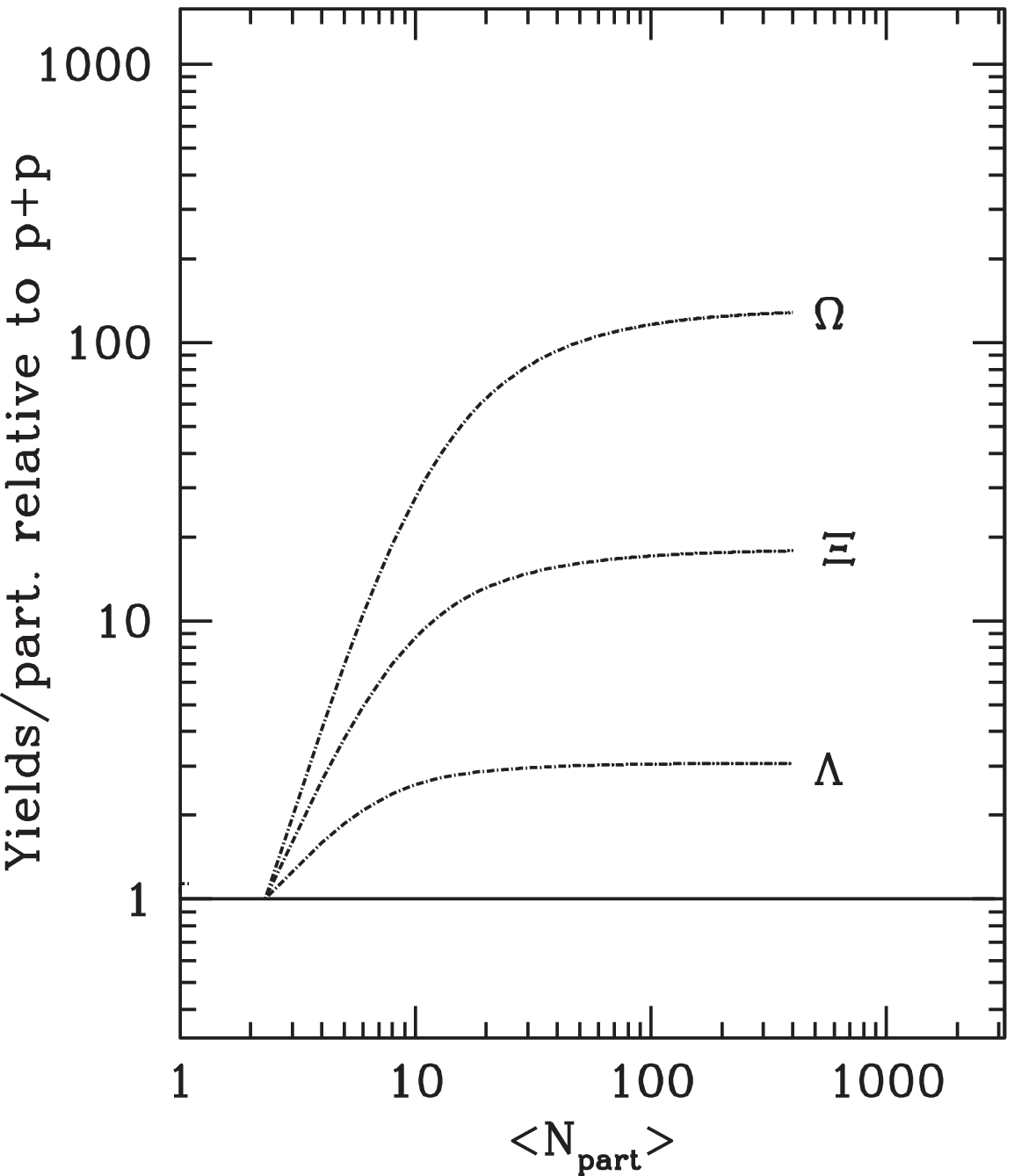}\\
  \caption{The calculated centrality dependence of strangeness ``enhancement" for different
  particles
  in Pb+Pb collisions at \sqrts= 8.73 GeV.~\cite{Redlich} }\label{Fig:CanonSuppres}
\end{center}
\end{minipage}
\begin{minipage}{0.46\linewidth}
    \begin{center}
  \includegraphics[width=\textwidth]{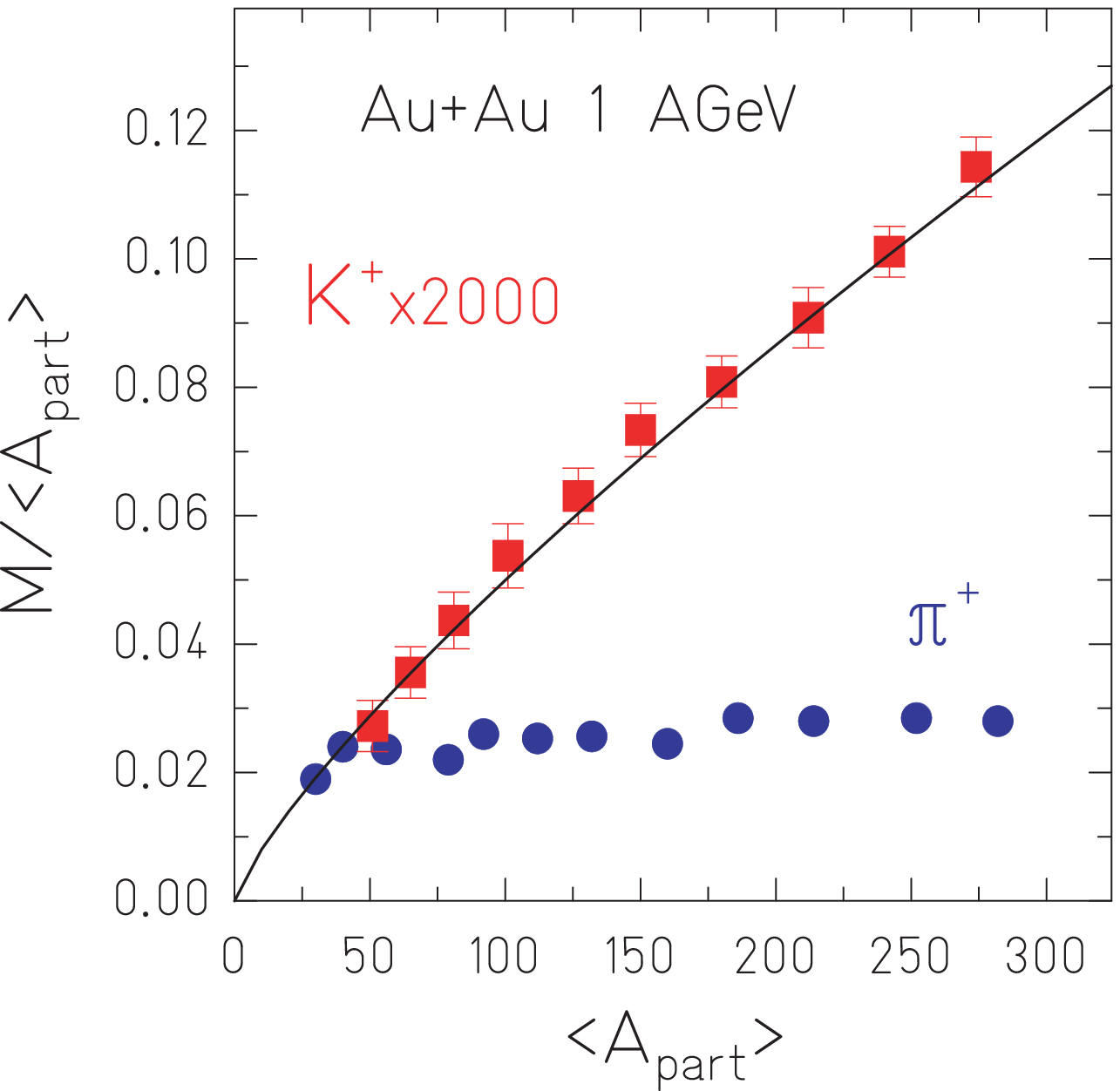}\\
  \caption{$K^{+}$ and $\pi^{+}$ multiplicities per participant as a function of $N_{part}$ for
  Au+Au collisions at \sqrts = 1 GeV\cite{Oeschler}.}\label{Fig:KaonEnhance}
  \end{center}
\end{minipage}
\end{figure}

 Statistical models utilize Grand Canonical Ensemble statistics,
which is only appropriate when the system becomes large.
In the Grand Canonical  approach quantum numbers need only be conserved on average. Thus one
can think of
creating an $\Omega^{-}$ without explicitly creating, at that moment, matching particles containing
 3 $\bar{s}$ quarks.  Ultimately all quantum numbers have to be conserved but at each step,
 when the Grand Canonical limit is reached, the system can temporarily ``pick up the slack".
 This leads to an interesting effect upon strangeness  production. In small systems,
 or the Canonical regime, all quantum numbers have to be conserved explicitly, this means there not only has to be
 energy available for strangeness creation but also the phase space. A small system  therefore
 results in a  suppression of strangeness, due to a lack of available phase space in which to
 create the quarks. Once the volume is sufficiently large this phase space suppression
 disappears and the amount of strange particle creation per unit volume becomes constant.
 The volume of the system is believed to be directly proportional to $N_{part}$. Although
 this is technically a suppression in smaller systems it is often referred to as an ``enhancement''
 in more central A+A collisions. The ``enhancement'' is measured experimentally as the
  yield per participant relative to the yield per participant in $p+p$ (or $p+Light$ nuclei when
  $p+p$  is not available). Figure \ref{Fig:CanonSuppres} shows the predicted behaviour
  for species as a function of volume, or $N_{part}$~\cite{Redlich}. As can be seen the larger the
  number of strange quarks in the particle the greater the phase space suppression effect. It has also been demonstrated that
  increasing the collision energy decreases the suppression of a given species.

\begin{figure}
    \begin{center}
  \includegraphics[width=0.8\textwidth]{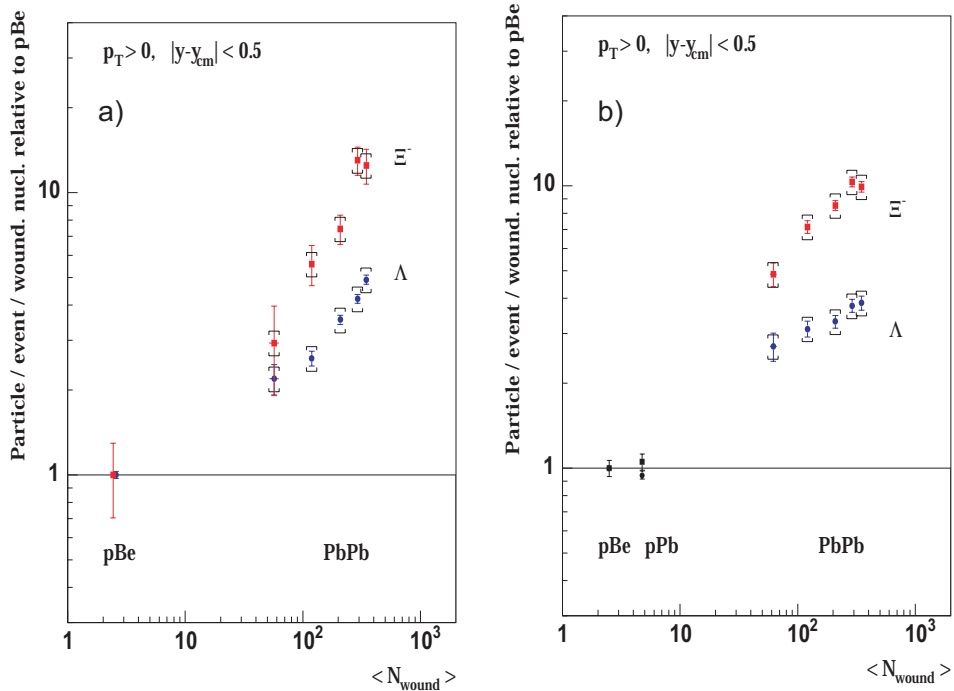}\\
  \caption{Hyperon enhancements as a function of number of wounded nucleons, $N_{part}$, for collisions at a)
    \sqrts = 8.8 GeV and b) \sqrts = 17.3 GeV as measured by the NA57 collaboration~\cite{Bruno}.
    Statistical and systematic errors are shown for each data point.}\label{Fig:NA57Enhance}
    \end{center}

\end{figure}

     At very low energies, such as those measured by the KAOS experiment at SIS, we can see the
affects of canonical suppression even in the kaons\cite{Oeschler}, Fig~\ref{Fig:KaonEnhance}.
However, as the collision energy increases this kaon suppression dissipates and it has been
shown that in Pb+Pb
collisions of \sqrts = 17.3 GeV even the $\Lambda$ yield per participant (or wounded nucleon) appears
to saturate, Fig.~\ref{Fig:NA57Enhance} b)~\cite{Bruno}. There is also the suggestion of a possible saturation
of the $\Xi$ yields in the most central data but the result is inconclusive. It would seem therefore
that the top energy CERN collision data are showing evidence of the applicability of the Grand
Canonical Ensemble for all particles up to the multi-strange baryons. The more recent
\sqrts = 8.8 GeV, (Fig.~\ref{Fig:NA57Enhance}a),  however, shows enhancement factors for the $\Xi$
and $\Lambda$ that are approximately
equal to the 17.3 GeV data. This result goes against our understanding of how canonical suppression
is related to collision energy. Calculations have shown that the enhancement for $\Xi$ should be
much higher at 8.8 than at 17.3 GeV. There are several possible explanations for this discrepancy.
One could be that the assumed linear relationship between correlation volumes and $N_{part}$ is incorrect. Another is that the temperature of the source reached in the lower energy
collision is not that assumed in the calculations, the enhancement factors being very sensitive
to this temperature.

\begin{figure}
\begin{minipage}{0.46\linewidth}
  \begin{center}
  \includegraphics[width=\linewidth]{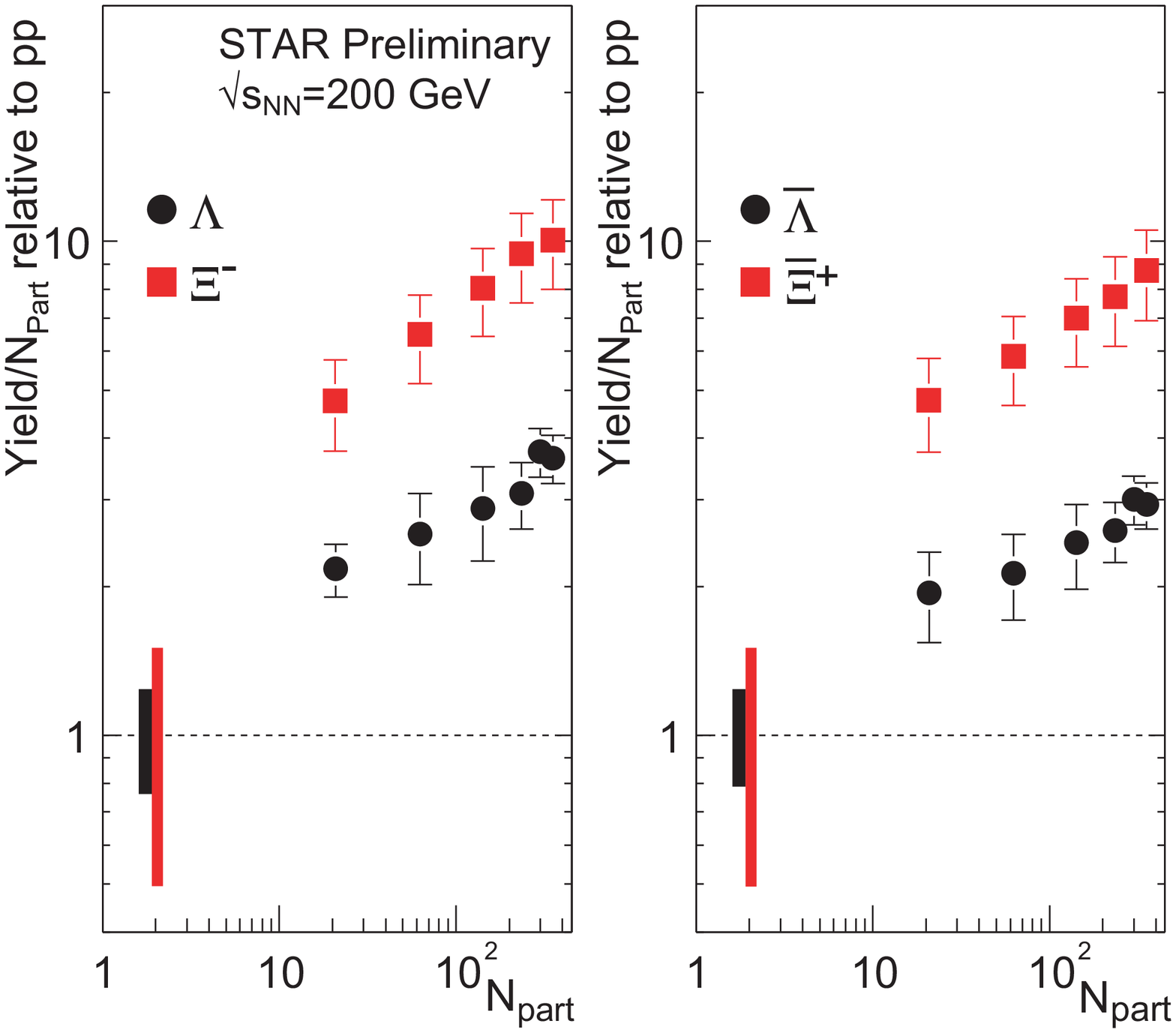}\\
  \caption{Enhancement factors as a function of $N_{part}$ for $\Lambda$ and $\Xi$. Error bars
  are statistical. The range for the $p+p$ results indicates the systematic uncertainty.}\label{Fig:STAREnhance}
\end{center}
\end{minipage}
\hspace{1cm}
\begin{minipage}{0.46\linewidth}
  \begin{center}
  \includegraphics[width=\linewidth]{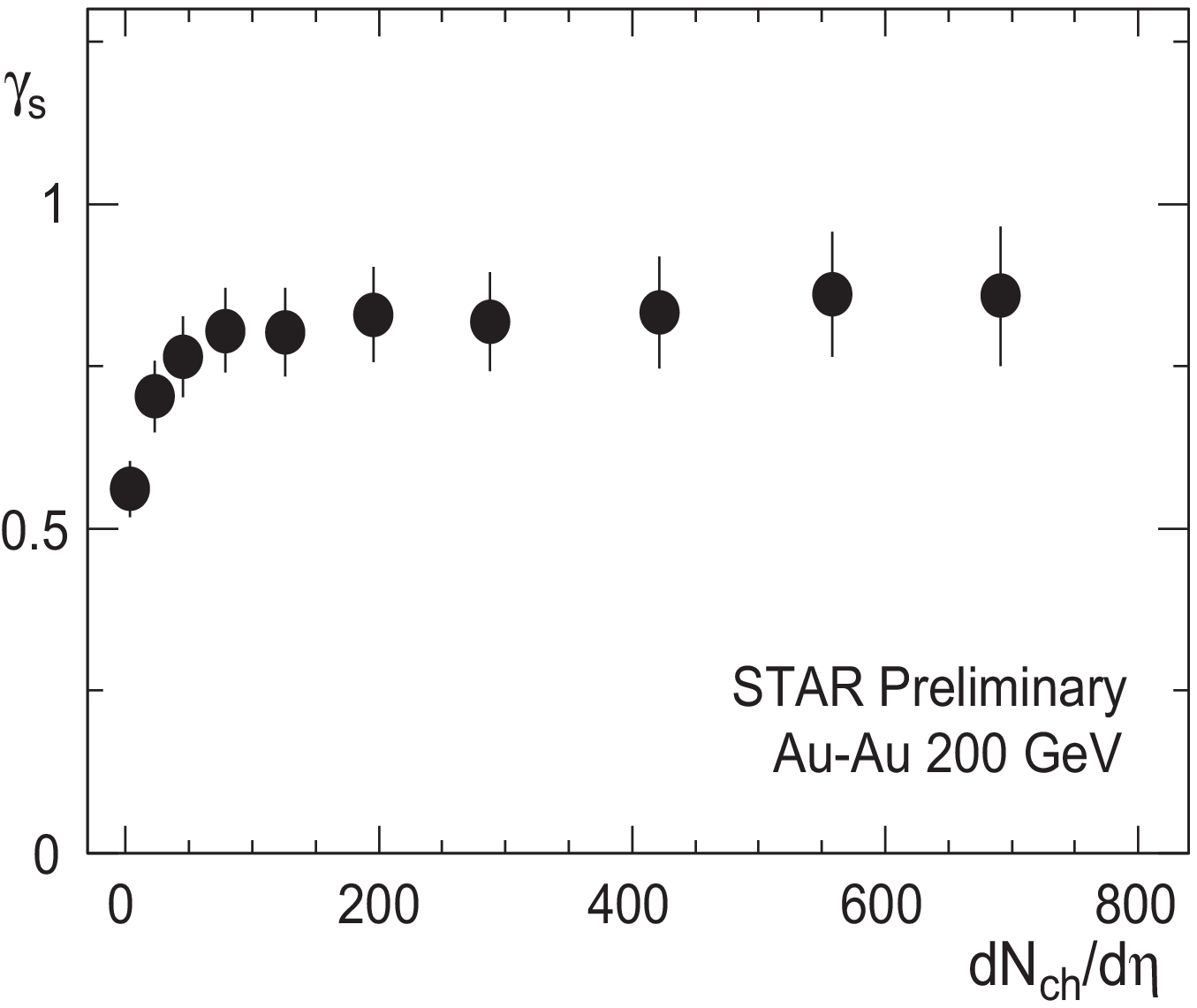}\\
  \caption{The $\gamma_{s}$ factor from statistical model fits of \cite{PBMStat} as a function
  of the number of charged particles for Au+Au collisions at \sqrts = 200 GeV.}\label{Fig:GammaS}
  \end{center}
  \end{minipage}
\end{figure}

     To try and gain further insight we turn to the RHIC measurements. Figure \ref{Fig:STAREnhance}
shows the preliminary measured enhancement factors for strange hyperons from STAR. We see that for
this data set the $\Lambda$ hyperons show no sign of reaching a plateau. As the SPS data appeared to saturate it is perhaps possible that the RHIC data shows an
over population of strangeness in the $\Lambda$ channel. However, Figure \ref{Fig:GammaS},
shows that the $\gamma_{s}$ factor only just reaches unity for the most
central data. This indicates that the system at RHIC is only just reaching
the Grand Canonical Ensemble limit for the most central collisions. So, once again, one is led
to the conclusion that the correlation volume is not simply proportional to $N_{part}$.
Further studies are needed to determine how the correlation volume can be mapped
, if at all, onto a physically measurable quantity.

     To conclude this section on system chemistry, we have established that for the most central
Au+Au collisions at RHIC statistical models can be applied and that the resulting fits reveal
physical quantities of the medium produced. It is also likely that at the top SPS energy and in the
more peripheral RHIC data the models can be used but that the interpretation should be made
with care.In elementary collisions and at lower energies the application of the
Grand Canonical Ensemble is likely not to be correct and whilst the data can be fit by
such models it is unlikely that the resulting parameters are related to physical source
temperatures and chemical potentials. As stated previously, statistical models do not try to
explain how the system came to be in equilibrium so the question "How did the system arrive
in this state?" remains. Microscopic hadronic models calculate that the system does not live
for a sufficient amount of time to achieve equilibrium through re-scattering in the hadronic phase.
An obvious mechanism is therefore to invoke a partonic phase where fast strangeness
equilibration is predicted. However, a transition to a deconfined phase has not yet been confirmed.

\section{Dynamics}\label{Section:Dynamics}

Having established the likely  properties of the source at chemical freeze-out we are next
interested in its dynamical properties. The time $\Delta t$  between T$_{ch}$ and T$_{fo}$,
Fig.~\ref{Fig:LightCone}, can have a significant effect on the
system. Elastic scatterings, while not changing the chemistry of the
system, may strongly affect the momentum spectra of the particles and much
of the transverse radial flow is built up during this
phase. Therefore I look into the question of collective motion, I discuss only evidence for
transverse radial flow as a whole session of this conference was dedicated to other types of
collective motion, see \cite{Flow}.

\subsection{m$_{T}$ Scaling}

\begin{figure}[htbp]
 \begin{minipage}{0.46\linewidth}
    \begin{center}
        \includegraphics[width=\linewidth]{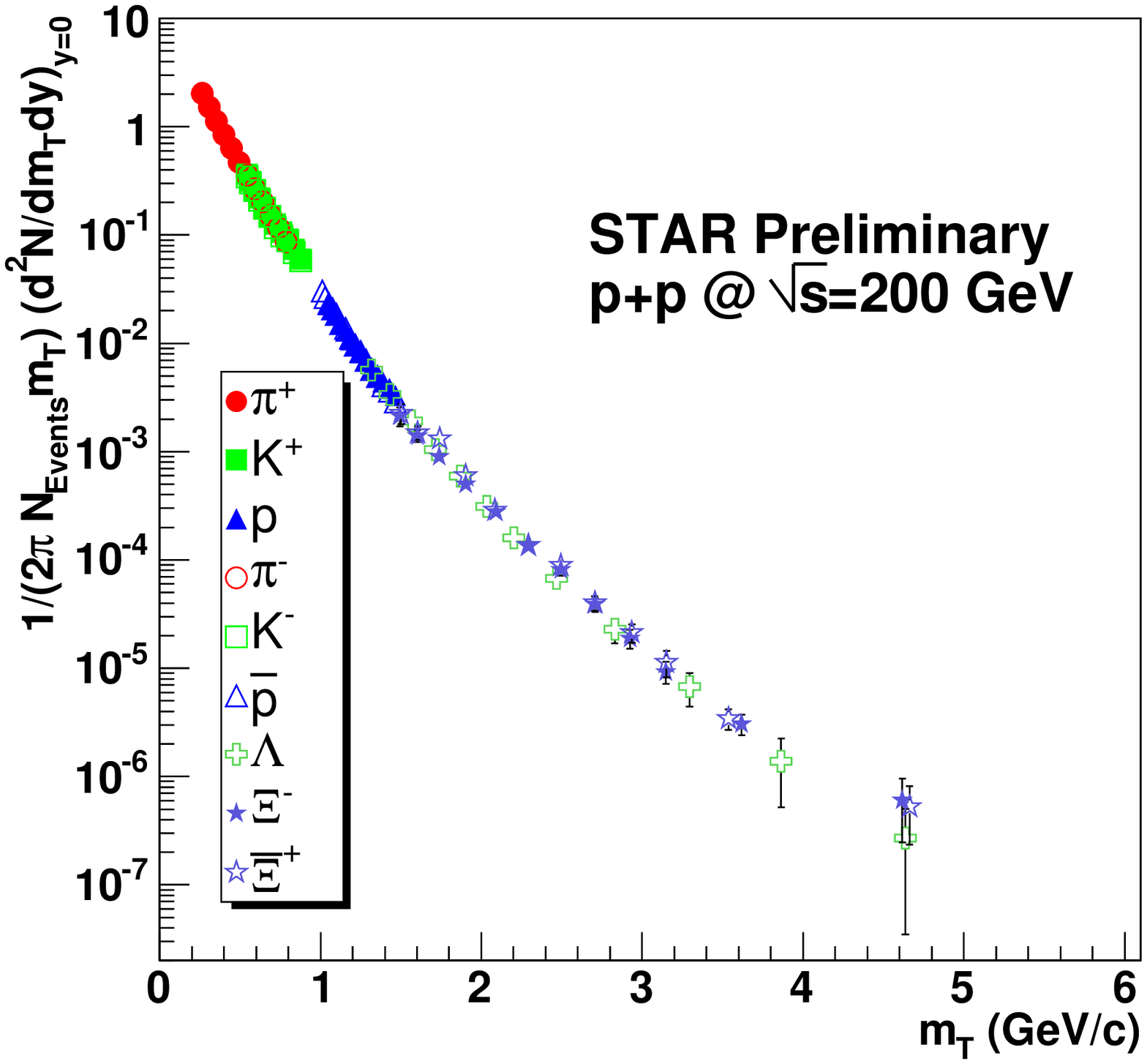}

\end{center}
    \caption{Preliminary $p+p$ spectra at \sqrts = 200 GeV from STAR. The data have
     been artificially scaled to get the best agreement with a universal curve.
     Errors are statistical only.}\label{Fig:ppScaling}

\end{minipage}
\hspace{1cm}
\begin{minipage}{0.46\linewidth}
    \begin{center}
        \vspace{-1cm}
        \includegraphics[width=\linewidth]{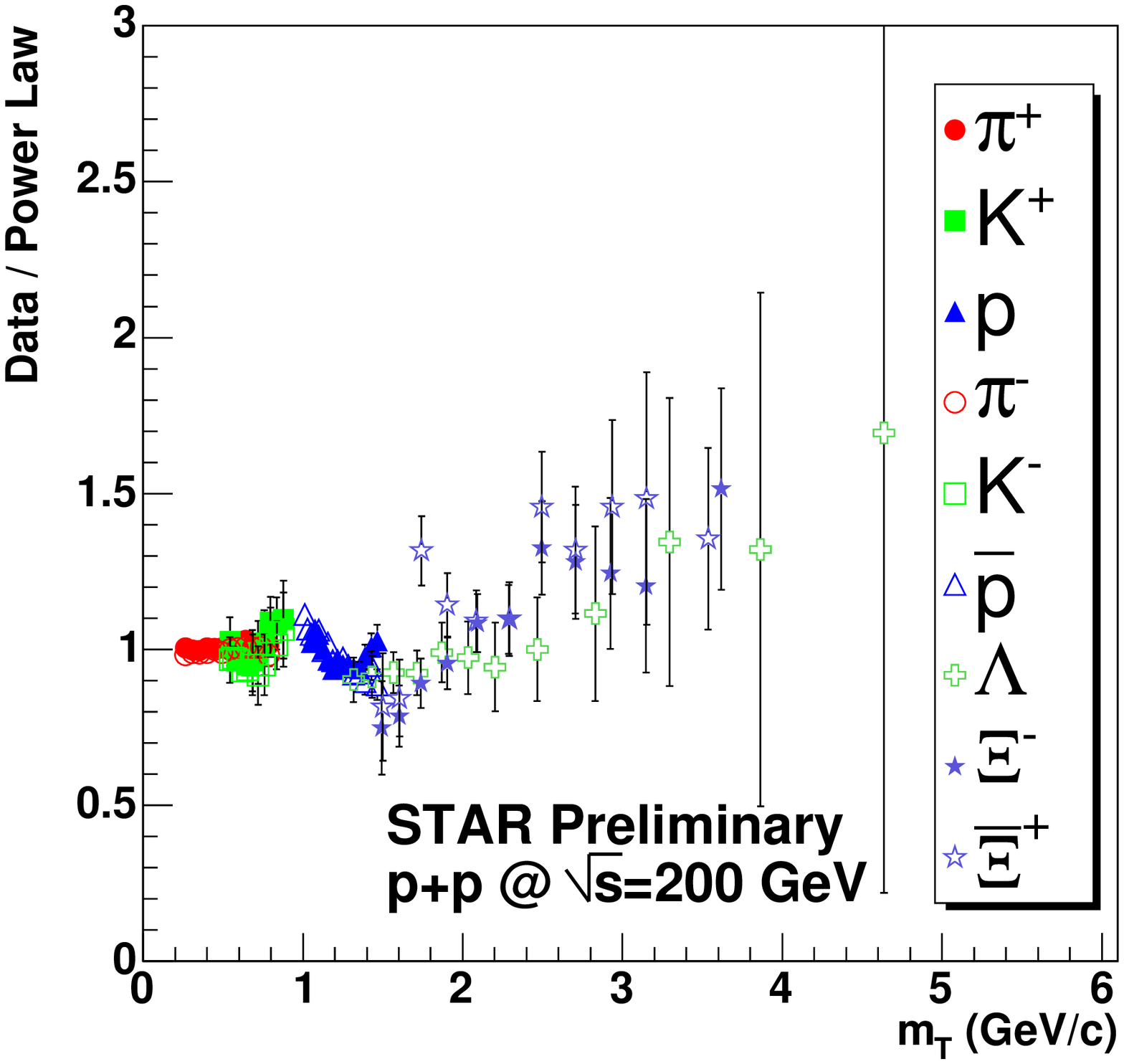}
    \caption{Ratio of data to a fit to a power law function for preliminary $p+p$ spectra from STAR
    after artificial scaling to get the best agreement between species.}\label{Fig:ppScaleRat}
    \end{center}
    \end{minipage}
\end{figure}

Before looking for evidence for transverse radial flow in heavy ion collisions
we first look at $p+p$ data from STAR. We wish  to see how the m$_{T}$ ( = $\sqrt{ (p_{T}^{2} + m^{2})}$)
spectra appear in a system where no radial flow is expected. Although we expect no collective
motion in elementary collision systems there is the possibility of m$_{T}$ scaling. The
phenomena of m$_{T}$ scaling means that the yields of all particles at a given m$_{T}$ are identical.
 Thus the m$_{T}$ spectra of all species will lie on a universal curve, ISR data have been
 successfully described in this manner~\cite{Dimitru}.
Fig.~\ref{Fig:ppScaling} shows the m$_{T}$ distributions for various particle species as
measured by the STAR collaboration in $p+p$ collisions at \sqrts = 200 GeV. The spectra
have been artificially normalized to obtain the agreement with a universal curve, so complete
m$_{T}$ scaling at 200 GeV is not observed. However, after scaling, the shape of the spectra
do appear to be universal, Fig.~\ref{Fig:ppScaleRat}.
For further discussion of the $p+p$ results at \sqrts = 200 GeV see \cite{Markpp}.

\begin{figure}
\begin{minipage}{0.46\linewidth}
  \begin{center}
  \vspace{-2.5cm}
        \includegraphics[width=\textwidth]{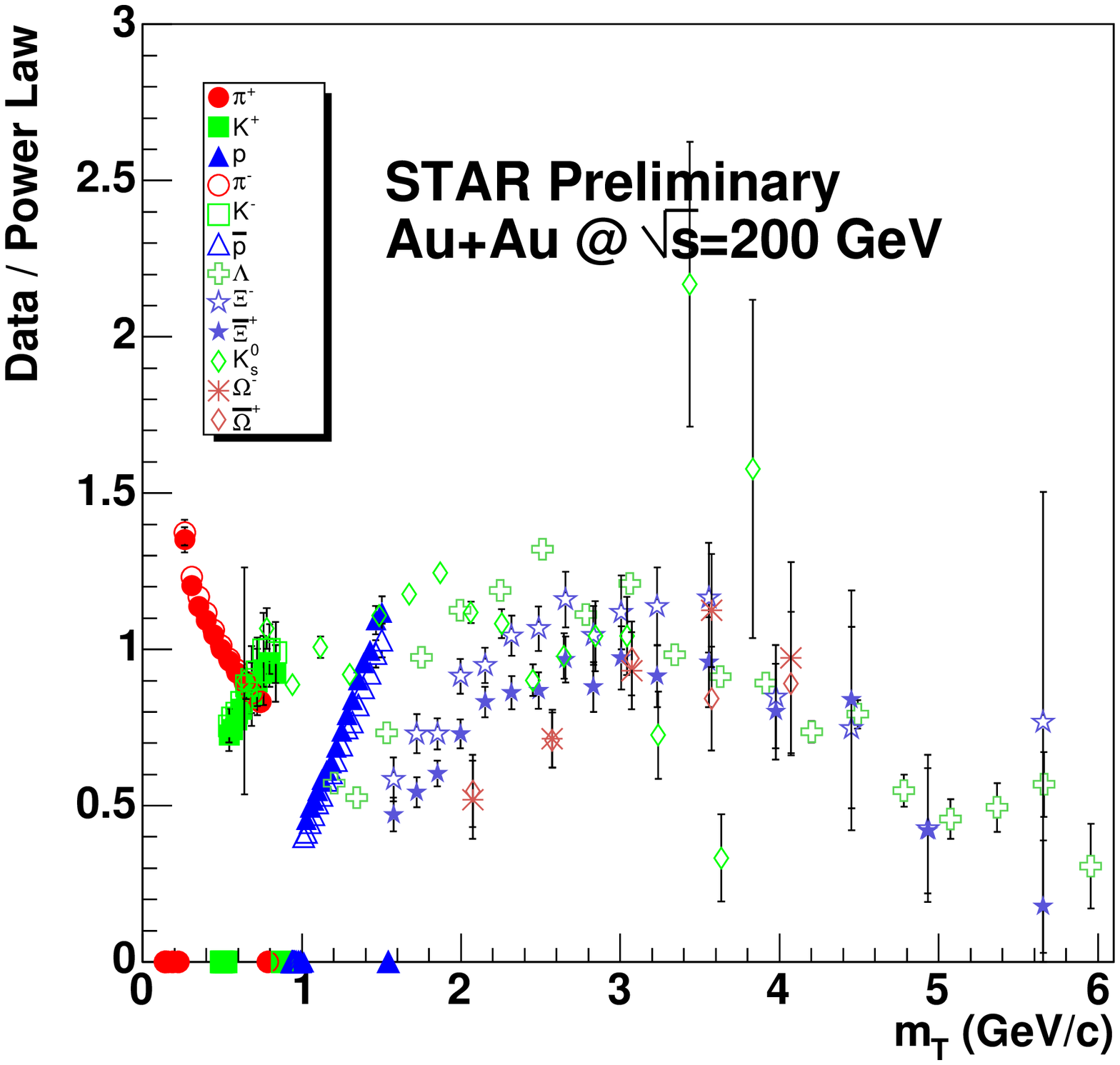}
    \caption{Ratio of data to a fit to a power law function for preliminary central Au+Au spectra
    from STAR
    after artificial scaling to get the best agreement between species.}\label{Fig:AuAuScaleRat}
\end{center}
\end{minipage}
\hspace{1cm}
\begin{minipage}{0.46\linewidth}
  \vspace{-2cm}
  \begin{center}
  \includegraphics[width=0.9\linewidth]{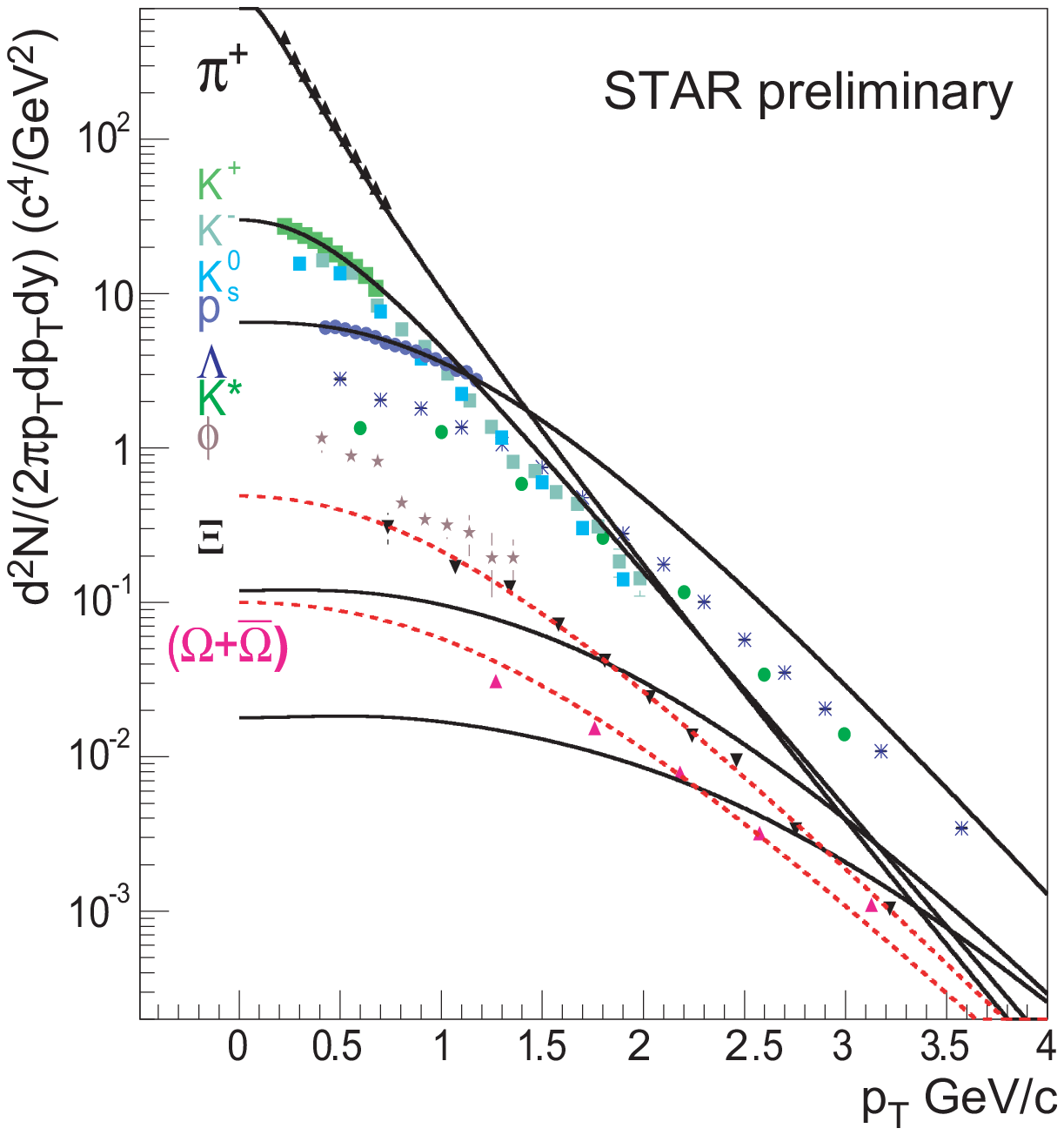}\\
   \end{center}
  \vspace{-1cm}
  \caption{ \pT spectra from top 5$ \%$  most central Au+Au data at
  \sqrts = 200 GeV. Solid curves are blastwave fits
  to $\pi$, $K$, and $p$. The dashed curves are fits to the $\Xi$ and $\Omega$.}
  \label{Fig:StarFlow}

  \end{minipage}
\end{figure}

    Even such an incomplete m$_{T}$ scaling is not applicable for Au+Au collisions,
as is evident in Fig.~\ref{Fig:AuAuScaleRat}. The Au+Au data show clear deviations for all
species at low m$_{T}$, evidence for radial flow, where the push from a common flow
velocity causes a depletion in all yields at low p$_{T}$.

\subsection{Transverse Radial Flow in Heavy-Ion Collisions}

To extract further information about the scale of the radial flow
in Au+Au collisions we try fitting a hydrodynamically inspired model, known
as the ``blastwave''~\cite{Blastwave}. This model assumes a common velocity profile
for all particles and that they freeze-out kinetically
with a common temperature, T$_{fo}$. The results of such fits
 can be seen in Fig.~\ref{Fig:StarFlow}. A common fit to the $\pi$, $K$, and $p$ spectra
yields a good representation of the data with T$_{fo}$ = 89 $\pm$ 10 MeV and
$< \beta >$ = 0.59 $\pm$ 0.05 c. The $\Xi$ and $\Omega$ spectra however
are much steeper, indicating a hotter freeze-out temperature. Indeed when a fit is made
to the $\Xi$ and $\Omega$ spectra alone, the dashed curve, T$_{fo}$ = 165 MeV $\pm$ 40 MeV and
 $< \beta >$ = 0.45 $\pm$ 0.1 c result. This suggests that the multi-strange baryons
 freeze-out thermally at an earlier time than the lighter particles.
 At this stage the source is hotter and the radial flow is lower; it has not yet had
 sufficient time to build up to its final value.

     The T$_{fo}$ of multi-strange baryons calculated from the blastwave
 model is very close to that calculated
 by the statistical models for T$_{ch}$. This is an indication that these
 particles decouple almost instantly upon hadronization, but  are already carring
 a significant radial flow. It is tempting to conclude that there is
 radial flow in the system before hadronization, $i.e.$ in a partonic phase.

     A slightly different interpretation comes from a complete
hydrodynamical simulation of the collisions~\cite{Kolb}. In this work the authors conclude
that the multi-strange baryons do not freeze-out at a significantly different time or with
a different radial flow velocity. However, the key point that there is an intrinsic radial
flow already , built up previous to hadronization is also one of their conclusions.

\section{Summary and Conclusions}

In summary I have shown that strange particle production and dynamics are key to understanding
the source created in heavy-ion collisions.

     The yields of strange baryons and mesons suggest a source that is in chemical equilibrium
for the most central A+A collisions and that at RHIC this source displays strangeness saturation.
The net-baryon number decreases smoothly with collision energy and is close to zero
at RHIC. The scale of the baryon transport has a strong effect on the rapidity distributions
of strange hyperons, which means that interpreting  mid-rapidity yields should be done
with caution. An enhancement of strange baryons per participant is observed in A+A collisions
when compared to $p+p$ at the same energy. This ``enhancement'' is most probably due to
a phase space suppression of the $p+p$ data. Obtaining such an enhancement requires a mechanism
for fast saturation of the strangeness phase space which hadronic transport models cannot provide
via the re-scattering of hadrons.

    The dynamics of the multi-strange particles suggest a sequential freeze-out after
hadronization that depends on the hadronic cross-sections. Thus the multi-strange particles
appear, in a blastwave scenario, to freeze-out kinetically very close to the chemical freeze-out
boundary. While this result relies on the validity of the blastwave approach both this
method and a full
hydrodynamical model calculate that the source created must have a sizeable radial flow during
the pre-hadronic stage.

    All these results, combined with those of the high \pT regime and v$_{2}$ measurements,
can be taken a strong evidence  of a system which, in the early phases, shows strong collective
motion, is very dense and has properties consistent with partonic degrees of freedom.

~


\begin{thebibliography}{99}
\bibitem{Markert:HotQuarks} C.~Markert, overview, "What can we learn from resonance production?"
These proceedings.
\end{thebibliography}
\end{document}